\documentclass[runningheads]{llncs}
\usepackage{graphicx}
\begin{document}
\title{Core Elements of Social Interaction for Constructive  Human-Robot Interaction\thanks{This work is part of the Hero project and is supported by the research program `Technology for Oncology' (grand number 15198), which is financed by the Netherlands Organization for Scientific Research (NWO), the Dutch Cancer Society (KWF Kankerbestrijding), the TKI Life Sciences \& Health, ASolutions, Brocacef, and the Cancer Health Coach. The research consortium consists of the Vrije Universiteit Amsterdam, Delft University of Technology, Princess M\a'axima Center, Centrum Wiskunde \& Informatica (CWI), and the University Medical Centers Amsterdam UMC.}}
%
%
\author{Mike E.U. Ligthart\inst{1}\orcidID{0000-0002-0768-9977} \and
Mark. A. Neerincx\inst{2,3}\orcidID{0000-0002-8161-5722} \and
Koen V. Hindriks\inst{1}\orcidID{0000-0002-8161-5722}}
\authorrunning{M.E.U. Ligthart et al.}
%
\institute{Vrije Universiteit Amsterdam, The Netherlands\\
\email{\{m.e.u.ligthart,k.v.hindriks\}@vu.nl}
\and Delft University of Technology
\email{m.a.neerincx@tudelft.nl}
\and TNO Soesterberg}
\maketitle              
\begin{abstract}
We present a discovery-based, first version, explicit model of social interaction that provides a basis for measuring the quality of interaction of a human user with a social robot. The two core elements of the social interaction model are \textit{engagement} and \textit{co-regulation}. Engagement emphasizes the \textit{qualitative nature} of social interaction and the fact that a user needs to be drawn into the interaction with the robot. Co-regulation emphasizes the \textit{interaction process} and the fact that a user and a robot need to be acting together. We argue that the quality of social interaction with a robot can be measured in terms of how efficiently engagement and co-regulation are established and maintained during the interaction and how satisfied the user is with the interaction. 


%

\end{abstract}
\section{Introduction}
Socially assistive robots (SARs) are developed to support people by means of social interaction~\cite{feil-seifer2005sar}. For example, we are developing a social robot (Hero) that aims to reduce stress among pediatric oncology patients. It makes sense that user studies that evaluate SARs focus on task performance (e.g. stress reduction). Success is often defined as a statistically significant difference between (mean performance scores of) the SAR condition and a control condition. Although essential, it does not paint the whole picture. There is typically a minority of participants for whom the interaction with the robot had no, or an opposite, effect. It is important to study these cases in order to make our socially assistive robots more inclusive.

In our studies we found several factors that influence the interpersonal differences in effectiveness. Factors that directly relate to the social interaction with a robot can have a significant impact on these differences. For example, the speed and intensity of the robot's speech and gestures~\cite{Ligthart2019getting} or speech recognition performance~\cite{ligthart2019long,Ligthart2020storytelling}. Other factors relate to user characteristics or (dynamic) contextual factors. For example, gender~\cite{ligthart2019long}
or too high expectations~\cite{Ligthart2017expectation}. Our aim in this paper is to provide a perspective on social interaction with a robot that allows us to focus more on these social factors. Although effectiveness and task performance are important, we argue that how the interaction with a social robot is shaped and experienced by users is just as important. We believe that by systematically studying quality measures derived from a model of social interaction we will also be able to improve the interaction designs of our social robots.

Initial steps to develop a model for social interaction have been taken in the context of the Hero project. A model that provides a basis for measuring the quality of interaction between a user with a social robot. It this paper we present the results of that initial discovery-based research. The remainder of this paper is organized as follows. In Section \ref{sec:def_social_interaction} we introduce the first version of the model and we discuss how it can be used to measure the quality of social interaction. In Section \ref{sec:case_study} illustrate how the hero project informed the model formation process.



%
%
%
\section{What is social interaction and how can it be measured?}
\label{sec:def_social_interaction}
Everybody has an intuitive understanding of what social interaction is. We engage in social interaction daily in numerous contexts. Examples of social interaction are taking a walk together, singing a duet, and having a conversation.\footnote{Bratman uses these three examples in \cite{Bratman2009modest} as examples of what he calls 'modest sociality'.} Social interaction is studied in many different research areas ranging from biology to philosophy, all taking a different perspective on what social interaction is exactly. We argue that an explicit model is required in order to be able to both quantitatively and qualitatively measure the quality of social interaction of a user with a robot. In this Section we first present such a model which we then use to establish how we can measure the quality of social interaction.

\subsection{Social Interaction}
Fong et al. \cite{fong2003survey} use the term \textit{socially interactive robots} to ``describe robots for which social interaction plays a key role''. Typically, in the literature on social robots, lists of (human social) characteristics that such a robot should have are provided. Good examples are again Fong et al.~\cite{fong2003survey} and more recently Breazeal et al.~\cite{breazeal2016social} who argue that a social robot should, for example, be able to express and recognize emotions, to communicate naturally using nonverbal and verbal language, have a personality, and more. Such lists are in line with Dautenahahn~\cite{dautenhahn1998art} who states that artificial social intelligence refers to an instantiation of human-style social intelligence in artificial agents. Presumably these characteristics contribute to and enable humans to conduct social interactions. Such lists however do not explicitly define what social interaction is and how it can be differentiated from other types of interaction and therefore do not provide a good basis for establishing how to measure the quality of interaction. Also, we believe that the emphasis on human social characteristics does not help to clarify when we can qualify the interaction between a human user and a robot as social.

In many works that try to distinguish social interaction from other types of (natural) interaction requirements are identified that set social interaction apart. Philosophers such as Margaret Gilbert, for example, argue that parties to a social interaction such as walking together simultaneously pursue a joint goal and to achieve this goal jointly depend on each other, which creates shared agency~\cite{Gilbert1990walking,gilbert2000sociality}. Social scientist James Coleman also emphasizes the interdependence of parties to a social interaction when he writes that ``a minimal basis for a social system of action is two actors, each having control over resources of interests to the other.'' Erving Goffman \cite{Goffman1959} emphasizes the reciprocal influence of individuals upon one another’s actions when in one another’s immediate physical presence, in the context of face-to-face interaction. Although these works identify essential elements that set social interaction apart from other types of interaction they do not provide much guidance on how to measure the quality of social interaction either.

\begin{figure}[!t]
\centering
\includegraphics[width=\textwidth]{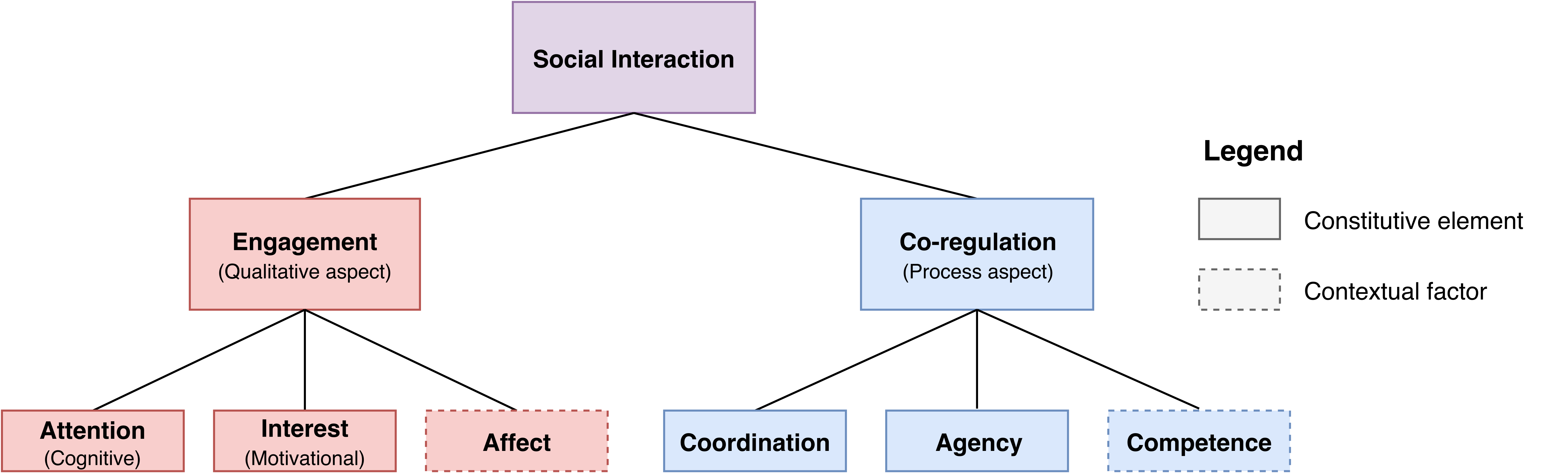}
\caption{First version of the model of social interaction}
\label{fig:social_interaction}
\vspace{-0.5em}
\end{figure}

An explicit model of social interaction should identify key components or the constitutive elements of what social interaction is. We use and base our own (simplified) definition on the work of De Jaegher, Di Paolo, and Gallagher \cite{DeJaegher2010social} who discuss and identify two such constitutive elements. According to De Jaegher, Di Paolo, and Gallagher, social interaction establishes a coupling between social actors that is engaging in nature ``as it starts to `takeover' and acquires a momentum of its own''~\cite{DeJaegher2010social}. Moreover, the shared agency of these actors must be co-regulated to maintain this coupling. Social interaction ends when this coupling is broken. For example, when the agency of one is inhibited by the other actor. Engagement and co-regulation are the two main components that need to be present to establish interaction that is social in nature between two or more actors (see Figure~\ref{fig:social_interaction}). Engagement is the qualitative aspect and co-regulation the process aspect of social interaction~\cite{DeJaegher2010social}. 
We further refine each of these two components of social interaction into two sub-components that constitute engagement and co-regulation and identify an additional contextual factor as a sub-component that varies the level of engagement and co-regulation. 

For social actors to engage in interaction, and especially to maintain engagement, they must have an interest in the interaction. Interest is one of the constitutive elements of engagement and refers to the motivational aspect of engagement. Engagement also has attention as a cognitive aspect~\cite{Corrigan2016engagement}. Attention is a constitutive element that directs engagement. Finally, engagement has an affective aspect as well~\cite{Corrigan2016engagement}. Affect is a contextual factor that does not constitute engagement but shapes the level of engagement. Affect influence engagement by either stimulating or inhibiting attention~\cite{Corrigan2016engagement} and interest~\cite{Morris2002affect}. For example, when people enjoy interacting with a robot more cognitive resources become available to pay attention to the robot~\cite{Corrigan2016engagement}. 

Co-regulation means that social actors coordinate the interaction they are engaged in together~\cite{DeJaegher2010social}. An example of such coordination is leaving pauses in a conversation to allow the conversational partner to take the turn~\cite{Garvey1981turntaking}. Social actors continuously and reciprocally influence each other's actions during the social interaction~\cite{Goffman1959} which requires agency on the part of each actor. It is important that during the interaction a sense of shared agency is maintained~\cite{Bratman2009modest,gilbert2000sociality}. In a social interaction there cannot be one actor that fully dictates how the interaction evolves. (Social) competences provide a contextual factor that influences the ability of each social actor to co-regulate the interaction~\cite{Dodge1985social}. For example, the more attuned to social signals these actors are, the more they can get in sync, the smoother the interaction proceeds~\cite{Gottman1975competence}.

%
\subsection{Quality}
An important aspect of the model is that it enables us to make predictions about what robot behavior contributes to a good social interaction. For each of the six elements that constitute or shape social interaction (see Figure \ref{fig:social_interaction}) we can establish levels and weigh how they contribute to the overall social interaction. Inspired by standard definitions of usability~\footnote{ISO standard for Ergonomics in human-system interaction (ISO 9241-210)}, we propose here to measure efficiency and satisfaction of social interaction using the six elements we identified. We argue that effectiveness which also is included in the usability standard is best measured in relation to task performance and is not an aspect of social interaction per se. 

%
%

\subsubsection{Efficiency}
The definition of social interaction discussed in the previous subsection gives us six elements (shown in figure~\ref{fig:social_interaction}) that either constitute or shape social interaction. The better each of the six elements of our definition are facilitated for the social actors the more efficiently a social interaction can be established. 

The robot influences the efficiency of the social interaction in two ways. It has a role in facilitating engagement and co-regulation for its human counterpart. For example, the robot must allow the interactor to have agency over the interaction as well. Secondly, the robot must appropriately communicate about its own attention, interest, affect, coordination, agency, and competence to allow people to sufficiently attribute these qualities to the robot. This allows people to perceive the robot as a social actor and experience the interaction as social~\cite{Wiese2017intentional}.

Practically, the efficiency can be measured by measuring how well each of the six social interaction elements are facilitated for the human and attributed to the robot. For example, measuring the willingness of participants to continue the interaction and their perception of the robot's willingness to continue is a way to measure interest.


\subsubsection{Satisfaction}
The final qualification of quality is the satisfaction people feel after the social interaction. Concretely, we propose to measure the subjective experience of the six social interaction elements. For example, how much agency did people experience or how competent did they feel talking to the robot?  

\section{Case Study: Hero Project}
\label{sec:case_study}
%
The model of social interaction and the approach we propose to measuring its quality provides a way to evaluate human-robot interaction more broadly. Although the definition is applicable for all socially assistive robots, the concrete measures and instruments that capture all the different facets of quality need to be tailored to the context of the robot application. For example, it matters which modality the robot uses to communicate. Coordination of the interaction is different when the user can talk to a robot or not. 

The research performed within the Hero project was part of the formation process of the model. In this section we collected a number of examples from the Hero project that illustrate the relevance of each element in the model and how it influences the quality of the social interaction between child and robot. In the Hero project a social robot companion is being developed for children with cancer. The goal is to reduce medical traumatic stress by accompanying the child throughout their time in the hospital. That is why we are working on facilitating a supportive long-term interaction~\cite{Ligthart2019getting,ligthart2019long} and on designing concrete stress reducing interventions. The case study includes two different user studies.

The first user study revolved around the child and the robot getting acquainted with one another. We evaluated five interaction design patterns that facilitate self-disclosure. The more comfortable the children feel to share about themselves, the more input the robot has to personalize future interactions~\cite{Ligthart2019getting} and the better a child-robot bond is facilitated~\cite{ligthart2019long}. The second user study was to evaluate a storytelling intervention. We evaluated three interaction design patterns that add different types of interaction to the storytelling activity~\cite{Ligthart2020storytelling}. 

\subsection{Attention}
Interacting with a robot without really giving it attention might not lead to the desired support it is meant to give. Measuring attention and use it as a measure of quality helps to evaluate robot behavior. For example, the more efficient the robot is able to draw attention to itself, the more useful it will likely be for a distractive intervention. We used eye gaze direction and a peripheral detection task to measure attention. A storytelling activity with and without the design patterns were compared. The story with patterns drew significantly more attention than the `plain' story~\cite{Ligthart2020storytelling}.

Looking at the efficiency helps to find an explanation of why the patterns were able to draw more attention. For example, during the plain story, where the robot continuously told the story, the attention dipped significantly half way. The design patterns introduced little moments of interaction that drew the attention back to the robot, preventing a dip~\cite{Ligthart2020storytelling}. In other words, the behaviors defined in the design patterns were more efficient in maintaining attention, than plainly telling a story.


\subsection{Interest}
That having and expressing a mutual interest in the interaction influences the quality of the interaction is something we encountered in the Hero project as well. Part of the getting acquainted interaction was a conversation about interests. 40\% of the children indicated that they shared interests with the robot. That group of participants was able to recall more about the robot's backstory (important for getting acquainted) than those who did not felt to have shared interests. And the more the robot expressed to have shared interests with the child, the more socially attractive the children rated the robot and the more they self-disclosed to the robot~\cite{Ligthart2019getting}. The easier children can relate to the robot, the better the social interaction quality, and the more children can benefit from the positive effects of interacting with the robot. Something that needs to be improved for the Hero robot. 

\subsection{Affect}
It is not only important to measure that users are engaged, but also whether they are engaged in a positive manner. Looking at the storytelling activity, almost all the children responded to the prompts of the robot to provide input for the story. In other words, a lot of engagement. However, only 59\% of the children expressed during the final interview to be satisfied with the prompts. A slight majority enjoyed the prompts, but some thought they were boring or distracting~\cite{Ligthart2020storytelling}. 

\subsection{Coordination}
The relevance for coordination during the interaction becomes abundantly clear when the robot operates autonomously. The main modality in the Hero project is speech. We therefore measure speech recognition performance and the influence the number of speech recognition errors has on other measures. Results show, for example, that the more speech recognition errors the robot made the less intimate children's self-disclosures were, the less comfortable the children felt during the interaction, the less socially attractive or competent the robot was rated, the less children could recall about the interaction, and the more confusing the interaction became~\cite{Ligthart2019getting,ligthart2019long,Ligthart2020storytelling}. 

\subsection{Agency}
One approach to keep an interaction manageable for an autonomous robot is for the robot to keep the initiative. However, it must do so without completely inhibiting the agency of its users. It can negatively affect the quality of the interaction otherwise. For example, a small portion of the children who engaged with the robot during the interactive story, by reenacting parts of the story with the robot (e.g. mimicking an elephant), expressed that they did so reluctantly. They felt a bit pressured by the robot or they did not want to disappoint it~\cite{Ligthart2020storytelling}. The better the robot can support children's agency, the higher the quality of the interaction. In this case the robot could, for example, communicate more clearly that it is perfectly fine to just watch the robot mimic an elephant.  

\subsection{Competence}
Social competence in human-robot interaction for the robot revolves around being able to perceive, process, and act on social signals of the user, while for people it is more about knowing how to navigate the robot's interfaces. The robot can help its user, for example, by offering an `how to talk to me'-tutorial. Children who interacted with the robot after a tutorial self-corrected more, for example by repeating an answer if they spoke before the speech recognition system was activated (indicated by a beep), preventing a lot of recognition errors~\cite{Ligthart2019getting}.

\section{Conclusion}\label{sec:conclusion}
In this paper we present a first model of social interaction that can be used to evaluate socially assistive robots. We distinguish between task performance and the effectiveness of the social interaction to offer `assistance' and the quality of the social interaction. We define the quality of social interaction in terms of the efficiency with which the social interaction is performed, and the satisfaction of a user with the ongoing social interaction. In order to arrive at a practical method for measuring the efficiency and satisfaction, we identified engagement and co-regulation as two key constitutive elements of social interaction. 

The exploratory work presented in this paper is discussed in the context of the Hero project, a socially assistive robot for children with cancer. A project that would benefit from a systematic evaluation of the social interaction quality using a mature model. The next steps would be to ground the model in the literature further and to validate it, hopefully collaboratively with the SAR community.

\bibliographystyle{splncs04}
\bibliography{references}

\end{document}